\documentclass[preprint]{vgtc}               

\graphicspath{{figures/}{pictures/}{images/}{./}} 

\usepackage{times}                     

\usepackage{tabu}                      
\usepackage{booktabs}                  
\usepackage{lipsum}                    
\usepackage{mwe}                       
\usepackage{amssymb}                   

\usepackage{graphicx}
\usepackage{tikz}
\usepackage{mathptmx} 

\usepackage{comment}
\onlineid{1011}

\vgtccategory{Research}

\vgtcinsertpkg

\title{Visualizing Uncertainty-to-Action Composition for Human Oversight}

\author{Chisom Anyabolu\thanks{e-mail: anyaboluc@rki.de}\\ %
        \scriptsize Robert Koch Institute %
\and Akshat Dubey\thanks{e-mail: dubeya@rki.de}\\ %
          \parbox{1.4in}{\scriptsize \centering Robert Koch Institute \\ Freie Universitat, Berlin} 
\and Georges Hattab\thanks{e-mail: hattabg@rki.de}\\ %
     \parbox{1.4in}{\scriptsize \centering Robert Koch Institute \\ Freie Universitat, Berlin}}

\teaser{
  \centering
  \includegraphics[width=\linewidth]{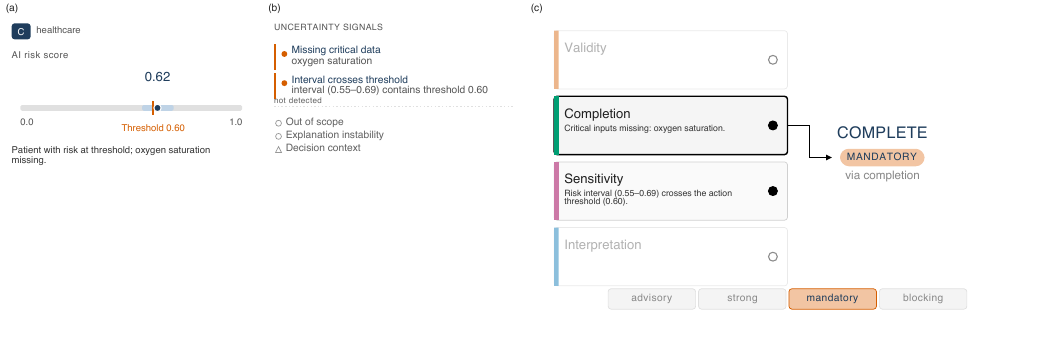}
  \caption{The ActionCue interface. (a) Case input, (b) detected uncertainty signals, and (c) the annotated precedence cascade. Completion and Sensitivity both fire; Completion wins precedence and binds to the oversight cue, \textsc{complete} at mandatory force, with Sensitivity retained as a supporting cue. Higher scores indicate greater predicted risk; the cue concerns the AI-supported decision path, not the predicted outcome.}
  \label{fig:actioncue-panels}
}
\abstract{
Artificial intelligence systems often disclose uncertainty, yet they rarely make clear what response that uncertainty should trigger.  
Most uncertainty visualizations encode uncertainty in model outputs, leaving users to discern the most appropriate course of action. A second region of the design space--uncertainty in the decision process itself, including how multiple uncertainty conditions compose into an oversight response-- remains comparatively underexplored.
We address this gap with two coupled contributions. First, we introduce an uncertainty-to-action binding framework that composes multiple uncertainty conditions into a single oversight response under a precedence policy with a contextual safety modifier. That response concerns whether and how an AI-supported decision may proceed, not the substantive domain decision itself. Second, we present ActionCue, a process-transparency visualization that renders that composition explicit. We demonstrate the approach through a three-way comparison with confidence-only and data-level uncertainty displays, using worked cases from healthcare, credit assessment, and disaster forecasting.
Together, the framework specifies how uncertainty conditions are resolved into an oversight response, and the visualization makes that resolution inspectable rather than implicit.
} 


\begin{document}



\maketitle

\section{Introduction}
\label{sec:introduction}

Most uncertainty visualization work has focused on encoding uncertainty in model output, using displays such as confidence intervals, distributional summaries, and related visual forms~\cite{padilla2018decision,hullman2018pursuit}. In accordance with Munzner's delineation between data abstraction and visual encoding~\cite{munzner2014visualization}, this places current existing work in one well-developed part of the design space: uncertainty is shown as something the user reads from the output, which is then interpreted to inform subsequent actions. Here, the interpretations depend on the individual's experience with AI, training background, and other factors~\cite{hattab2024persona}.
AI systems are increasingly transparent about uncertainty in this sense; however, they rarely make clear what response that uncertainty should trigger.

A second part of the design space remains underexplored: the degree of uncertainty inherent in the decision-making process itself. This includes which conditions are present, how they compose, and what oversight response that composition supports. This matters because the problem is often not that uncertainty is hidden, but that the link from uncertainty to action is missing or opaque. As a result, users are required to compose the response themselves, and that composition is difficult to inspect, compare, or audit~\cite{alfrink2023contestable}.

Consider a clinical risk model used to assess the probability of adverse outcomes in healthcare settings. The model reports a score against an action threshold, together with a confidence interval around that score. Suppose that a required input is missing and that the interval also crosses the threshold, so the score cannot be resolved against it. 
The interface makes uncertainty visible; however, it does not provide guidance on whether the clinician should proceed, complete the missing information, reassess, or escalate. The issue extends beyond how to convey uncertainty in an estimate: it is how to represent the process by which multiple uncertainty conditions resolve into an oversight response.

We address this gap with two coupled contributions. First, we introduce an uncertainty-to-action binding framework whose central move is deterministic composition: when multiple heterogeneous uncertainty conditions are present at once, a precedence policy with a contextual safety modifier resolves them into a single oversight response. 
That response governs whether and how the AI-supported decision may proceed; the framework does not resolve the substantive domain decision itself.
Existing approaches bind a single uncertainty condition to a response; none specify how several conditions compose. Second, we present ActionCue, a process-transparency visualization that makes the composition explicit. 

\section{Background and adjacent work}
\label{sec:background-adjacent-work}

A number of adjacent literatures have formalized partial relations between AI uncertainty and human response. Selective prediction is grounded on the premise that model confidence is contingent on abstention when confidence falls below a predetermined threshold~\cite{geifman2017selective}. Appropriate reliance studies how interfaces calibrate users' trust, ensuring that they adhere to accurate predictions and override them when erroneous~\cite{lee2004trust,bansal2021does,buccinca2021trust}. A communication perspective on AI trust models how trustworthiness cues are conveyed and processed into trust judgments~\cite{liao2022designing}. Tiered clinical decision support binds alert severity to workflow interruption, with more consequential alerts interrupting users more forcefully~\cite{phansalkar2013drug}. Counterfactual recourse binds a model output to possible input changes that could alter the decision~\cite{wachter2017counterfactual,ustun2019actionable}. XAI question banks bind user questions regarding model behavior to explanation types that can answer them~\cite{liao2020questioning}.

The significance of these literatures lies in their demonstration that uncertainty is not merely a passive phenomenon; rather, it can be associated with action, interruption, explanation, or recourse. However, each focuses on a single aspect of the broader uncertainty-to-action relation, and none provides a shared schema for cases in which multiple heterogeneous conditions are present, such as an out-of-scope input, a missing data field, and a borderline interval. In such cases, the user is left to compose the response without assistance, and the link from detected uncertainty to action remains opaque.

The field of uncertainty visualization has developed a rich vocabulary for representing uncertainty as a property of data or model output. Various data visualization techniques, including confidence intervals, distributional summaries, gradient hulls, fan plots, ensemble overlays, hypothetical outcome plots, and quantile dotplots, have been employed to encode uncertainty around a quantity, forecast, or decision-relevant estimate~\cite{padilla2018decision,hullman2018pursuit,maceachren2005visualizing}. Recent studies have examined how such encodings support probabilistic decision-making~\cite{kale2019decision}, how uncertainty communication varies across visual, textual, and spoken modalities~\cite{stokes2024voicing}, how multiple-forecast displays compare for decision tasks~\cite{matzen2024effects}, and how diagnostic uncertainty can be encoded in tabular settings~\cite{yener2025visualizing}.

This series of studies is characterized by a shared abstraction-level choice: uncertainty is rendered as a property of the output, and the user interprets that output to determine how to make an informed decision. Our paper addresses a complementary abstraction: uncertainty in the decision process. Our focus is not only on the presence of uncertainty; it is also on how multiple conditions are organized, resolved, and bound to an oversight cue through informed heuristics integrated during the design process.

\section{Design framework}
\label{sec:framework}
The framework operationalizes uncertainty-to-action binding through five elements: the uncertainty source, the decision context, the action family, the responsible actor, and the workflow force. 
The uncertainty source is defined as what has become uncertain or unreliable in the AI-supported decision, such as missing input data, model scope violation, threshold sensitivity, or explanation disagreement. The decision context names properties that influence the requisite response, encompassing harm, reversibility, and accountability. The action family delineates the tasks to be performed; the responsible actor identifies the individuals tasked with execution; and the workflow force determines the extent to which the interface should intervene in the user's task flow. The underlying signals may be continuous or probabilistic. Scope violation, explanation disagreement, and proximity to a threshold are graded quantities. Composition therefore operates on deployment-defined predicates that determine when such a signal constitutes an oversight-relevant condition.

These elements are not treated as freely combinable dimensions. They are constrained by a predetermined rule table, thereby establishing a structured framework. The rule table adapts established binding between AI uncertainty and human response into a single structured form, while the specific rule entries are stipulated to be refined through domain expertise and evaluation. Each rule has a condition and a consequence: when a given uncertainty source is present in a decision context, the rule assigns a base action, force level, responsible actor, and rationale. In the event that multiple rules are triggered, the composition policy consolidates them into a unified primary oversight cue, supplemented, when pertinent, by a supporting cue stack.

We define four rule classes. \textit{Validity rules} are invoked when the AI output falls outside its intended scope or is unsuitable for its intended application~\cite{geifman2017selective}. The \textit{completion rules} pertain when a required input is missing or unreliable. \textit{Sensitivity rules} are initiated when the uncertainty interval intersects or approaches the action threshold. \textit{Interpretation rules} fire when explanations are unstable or models disagree~\cite{alvarez2018robustness}. These classes provide a minimal taxonomy for composing uncertainty-to-action rules; additional classes may be added in domain-specific deployments.

The framework defines six action families, namely, \textit{proceed}, continue with the AI-supported action; \textit{inspect}, examine the uncertainty before acting; \textit{complete}, collect, verify, or repair required information; \textit{reassess}, delay, monitor, or rerun the decision after new information arrives; \textit{escalate}, transfer to a higher expertise or authority level; and \textit{abstain}, withhold the AI recommendation or block AI-supported action. These families govern use of the AI-supported decision path; they do not prescribe the substantive domain decision. A cue to complete or escalate concerns whether the AI-supported decision is fit to proceed, not the diagnosis, credit determination, or evacuation decision itself. We treat documentation as an accountability requirement that may attach to any action, rather than as a separate action family.

Workflow force is defined as the degree to which the interface intervenes in the decision flow. We adapt a four-tiered vocabulary from tiered clinical decision support~\cite{phansalkar2013drug}. \textit{Advisory} cues are non-interruptive and informational. \textit{Strong} cues are non-interruptive but recommend a direction. \textit{Mandatory} cues interrupt task progression until the user acknowledges, acts, or explicitly overrides the cue. \textit{Blocking} cues impede the execution of the AI-supported action or withhold the recommendation until the underlying condition is addressed and resolved. 
The action family and the workflow force are not fully independent. Table~\ref{tab:constraint_matrix} specifies the pairings permitted in the current instantiation. These are design stipulations motivated by coherence between what an action demands and how forcefully an interface should intervene; they require domain-specific validation and may be revised.

\begin{table}[tb]
  \caption{Permitted pairings of action family and workflow force. \checkmark{} marks pairings that a base rule may assign directly. $\dagger$ marks pairings not assignable by a base rule but reachable when the safety modifier elevates force by one level. Empty cells are excluded by the framework.}
  \label{tab:constraint_matrix}
  \scriptsize%
  \centering%
  \begin{tabu}{%
    l%
    *{4}{c}%
    }
  \toprule
   & \multicolumn{4}{c}{Workflow force} \\
  \cmidrule(lr){2-5}
  Action family & Advisory & Strong & Mandatory & Blocking \\
  \midrule
  Proceed  & \checkmark &            &            &            \\
  Inspect  & \checkmark & \checkmark & \checkmark & $\dagger$      \\
  Complete &            & \checkmark & \checkmark & \checkmark \\
  Reassess &            & \checkmark & \checkmark & $\dagger$       \\
  Escalate &            & \checkmark & \checkmark & \checkmark \\
  Abstain  &            &            & \checkmark & \checkmark \\
  \bottomrule
  \end{tabu}%
\end{table}

It is important to note that a single case may be subject to the influence of multiple rules. The framework addresses this through a deterministic precedence policy, wherein validity takes precedence over completion, completion takes precedence over sensitivity, and sensitivity takes precedence over interpretation. The highest-precedence firing class establishes the primary cue, encompassing the action, actor, force, and rationale. Lower-precedence firings are retained as supporting cues when compatible. In the absence of a rule that is triggered, the framework returns \textit{proceed} with advisory force.

The precedence order reflects an epistemic dependency rather than a priority preference. Each class presupposes that the classes above it are satisfied for its own signal to carry meaning. A model applied outside its intended scope still yields a score and an interval; however, these quantities are not yet trustworthy, so a sensitivity reading taken from them is uninformative. The same dependency holds one level down: when a required input is missing, the interval is computed on incomplete data, and its position relative to the threshold is not yet a reliable signal. Sensitivity, therefore, becomes meaningful only once validity and completion are established. Completion winning precedence over sensitivity does not demote a genuine threshold risk; it defers the sensitivity reading until the input that would make it trustworthy is present. We treat a required input as one whose absence degrades the downstream metric, not merely any field absent from the schema. High-stakes danger is not lost under this ordering. It is absorbed on a separate axis: the safety modifier raises the workflow force of the resolved cue, so a hazardous case is rendered as a higher-force response rather than a reordered one.

The policy resolves precedence between classes but not plurality within a class; when multiple rules of the same class fire, a within-class ordering would be required. A natural extension is to rank co-class rules by a severity weight, so that the most consequential firing sets the primary cue while the others are retained as supporting cues. We leave this to future work and treat each class as contributing a single resolved firing in the present framework.

Harm and irreversibility function as a safety modifier rather than a rule class. When the decision context entails significant potential for harm or limited reversibility, the modifier elevates the workflow force of the resolved cue by one level: advisory becomes strong, strong becomes mandatory, and mandatory becomes blocking. It applies only to a cue produced by a fired rule, and it raises force without altering the action. This maintains the distinction between uncertainty sources, which initiate rules and determine the action, and the decision context, which determines how forcefully the interface intervenes.

We deliberately decouple action from force. The action family answers what the uncertainty condition demands and is fixed by the uncertainty source: a missing input demands completion, a fragile interval demands reassessment, whatever the stakes. Harm does not change what is epistemically wrong with the decision, only how forcefully the interface should respond, so the two are orthogonal. Coupling them would require a separate action mapping for each combination of uncertainty source and context, and would reintroduce the non-determinism the framework exists to remove. High stakes are met at the top of the scale: a reassess or inspect cue elevated to blocking withholds execution until the condition is addressed, preserving the rationale a bare abstain would discard. Force elevation does not reassign the responsible actor; that element is fixed by the firing rule, so harm-driven escalation to a higher authority is a separate extension rather than a modifier effect. 

\section{ActionCue: visualizing uncertainty-to-action binding}
\label{sec:actioncue}

ActionCue is a three-panel Streamlit prototype built over a small set of synthetic decision cases. Each case stipulates its uncertainty signals rather than estimating them from a model, allowing us to isolate the contribution of the prototype: the binding between uncertainty conditions and oversight actions, rather than upstream predictive accuracy. As shown in \cref{fig:actioncue-panels}, the interface displays the case input, encompassing the risk score and the interval in relation to the action threshold; the detected uncertainty signals; the composition trace; and the resulting oversight cue.

The central visual idiom employed is an annotated precedence cascade. In contrast to uncertainty displays that primarily encode uncertainty at the data level, such as intervals or ensembles, ActionCue encodes uncertainty at the process level. This entails the following: which rule classes are active, how they compose under precedence, and what response the composition produces. Precedence is encoded by vertical position: validity precedes completion, completion precedes sensitivity, and sensitivity precedes interpretation. The rule class is encoded by band position and categorical color. The firing status is encoded by fill, while the winning class is marked by border emphasis.
These channel assignments follow effectiveness rankings for the underlying data types~\cite{munzner2014visualization}. Precedence is an ordinal attribute, an ordered sequence of discrete classes rather than a continuous quantity, and spatial position is the most effective channel for ordered data. Discrete banding and categorical color reinforce that the vertical axis encodes order rather than magnitude, so it is not read as a continuous or quantitative scale.
A labeled resolution arrow connects the winning class to the output cue, thereby rendering the composition policy visible rather than implicit. The workflow force scale is shown at the bottom, with the assigned level marked. In addition, a rule registry tab shows the four rules that are read directly by the rule engine, without an intermediate state.

The foreground encoding remains at the class level. Information regarding the particular rule that was activated, the triggering signal, and the underlying rationale for the rule are available on demand through hover or expansion~\cite{shneiderman1996eyes}. This separation is intentional. The current framework delineates precedence between classes; however, it does not yet address within-class plurality or tie-breaking. Consequently, foregrounding rule-level cells would imply a resolution granularity that the framework does not yet support. The code and data for ActionCue are publicly available at \url{https://github.com/Sombiri/actioncue}. A deployed prototype is available at \url{https://actioncue.streamlit.app/}.


\subsection{Comparison across displays}
\label{sec:three-way-comparison}

\begin{figure}[ht]\centering
\includegraphics[width=.50\textwidth]{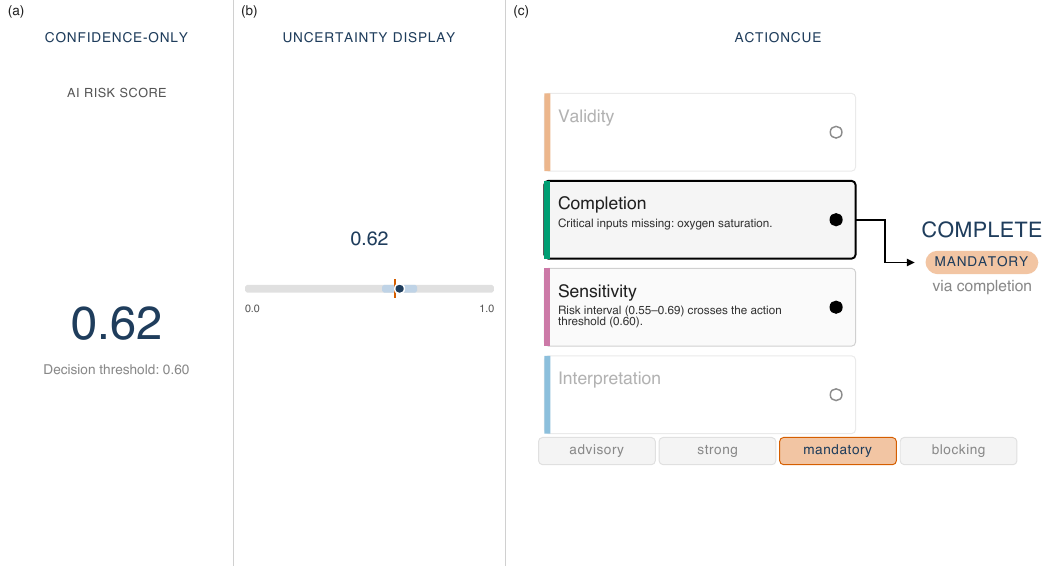}
\caption{Case~C under three displays. (a) Confidence-only supports acting on the point estimate; (b) the uncertainty display marks the case borderline via the interval; both omit the missing oxygen saturation. (c) ActionCue composes the conditions---Completion wins precedence over Sensitivity---and surfaces the missing input as the primary oversight cue (\textsc{complete}, mandatory force).} 
\label{fig:three-way-comparison}
\end{figure}

\cref{fig:three-way-comparison} compares a single clinical case under three displays. Case~C has a risk score of 0.62 against an action threshold of 0.60, with a risk interval of 0.55 to 0.69, and there are two uncertainty conditions: oxygen saturation is missing and the interval crosses the threshold.

A confidence-only display shows the point estimate against the threshold. Because 0.62 exceeds 0.60, the display presents the case as one in which the AI-supported action is indicated. It does not expose the missing input, the interval crossing, or any reason to question the recommendation.

A conventional uncertainty display adds the interval. Because the interval spans the threshold, the case appears borderline. This improves the display by showing uncertainty in the estimate; however, it still cannot represent a missing required input. Oxygen saturation is not an uncertain value in this case; it is absent. The most actionable signal therefore remains outside the display.

ActionCue renders the composition itself. The cascade shows that completion and sensitivity both fire. Completion wins precedence, producing the cue to complete the missing oxygen saturation before acting, at mandatory force. Sensitivity remains visible as a supporting cue. The interval crossing, which the uncertainty display foregrounds, is retained as a supporting cue. The missing input, which the previous displays cannot represent, becomes the primary actionable issue.

Across the three displays, ActionCue does not merely add more information. It changes the decision object from an uncertain estimate to a resolved oversight response. The contribution is therefore not only visual encoding, but the explicit representation of how multiple uncertainty conditions are ordered, resolved, and bound to action.

\subsection{Generativity across cases}
\label{sec:generativity-across-cases}

The same composition machinery applies across domains. We illustrate this with cases from healthcare, credit, and disaster forecasting. Case A, a complete, in-scope clinical case whose risk interval does not cross the action threshold, fires no rule and produces a proceed cue at advisory force. Case~CR, a credit decision, fires validity and completion; validity wins precedence and produces an escalate cue at mandatory force, with completion retained as a supporting cue. Case~DR, a disaster forecasting decision, fires sensitivity and produces an inspect cue at strong force. Because the decision context has high harm and low reversibility, the safety modifier raises the force to mandatory.


\section{Discussion and limitations}
\label{sec:discussion}
ActionCue targets a \emph{domain translator}: a decision-maker who is accountable for acting on model outputs and can reason about procedural conditions and their implications, but is not expected to inspect model internals or feature-level attributions. Accordingly, the interface operates at the level of process oversight: it communicates which uncertainty conditions are present and what response they compose, rather than exposing the model internals that produced them. This scopes the intended evaluation to whether such users can read and act on the composed cue, and treats data-level model explanation as a separate concern.

The framework and visualization have two main implications. First, ActionCue situates decision-process uncertainty within a distinct region of the uncertainty-visualization design space, separate from the data-level encodings that predominate in prior work. Data-level encodings represent uncertainty about a model's output, whereas the composition trace represents uncertainty in the decision process that leads to an oversight response. In this regard, the contribution is not merely a new display, but a way to make the binding between detected uncertainty and an accountable response visible.
Second, situating uncertainty at the process level relocates interpretation rather than removing it. The user no longer derives the oversight response unaided from an uncertain quantity; instead the triggering conditions, their precedence, and the resulting response are made inspectable, and the cue can be contested on the specific condition that produced it. The shift is nonetheless consequential: a mandatory cue interrupts progression and a blocking cue withholds AI support until the condition is resolved, so the framework constrains the decision path even as it makes that constraint legible. The separation between oversight response and substantive domain decision holds only where constraining or withholding AI support does not itself determine the latter; settings in which the AI is the sole decision procedure fall outside the present scope.

It is important to note that the scope of the prototype is inherently limited. First, we stipulate uncertainty signals rather than estimating them, so coupling the framework with live uncertainty estimators remains future work. 
The displayed interval should therefore be read as a provisional model output used to demonstrate composition, not as a calibrated estimate under a missing-data mechanism. How missingness should propagate into an interval depends on the estimator and the missing-data treatment, neither of which the prototype models.
Second, our comparison uses a small set of foils rather than a comprehensive survey of uncertainty idioms. This scope reflects our focus on the representational capacity of data-level displays to illuminate the decision process, rather than the relative merits of specific visual styles. While more elaborate idioms such as hypothetical-outcome plots~\cite{hullman2015hypothetical} and ensemble displays differ in visual encoding, they do not differ in whether they represent composition in the decision process. 
Third, the framework does not yet fully specify rule-level plurality; same-precedence ties and within-class rule resolution remain open, as does a more nuanced decomposition of explanation-side uncertainty~\cite{dubey2026ubiqtree}. Fourth, composition operates over detected conditions. Where the magnitude of a signal, or the uncertainty attaching to it, must itself be jointly reasoned about to select the oversight response, the present rule-based composition is insufficient.

Consequently, future research is structured around these limits. The most immediate is empirical evaluation of the composition trace as a visual encoding, whose primary outcome is oversight appropriateness: whether the accountable decision-maker reaches the warranted response more reliably than with confidence-only or data-level displays. A second concern the rules themselves. The predicates and rule mappings are deployment-specific policy choices requiring domain expertise, evaluation, and versioning. Making them explicit relocates part of the accountability for an oversight response from the individual operator to the rule set and those who maintain it. We treat this as a property of the design rather than an unintended effect: a composition performed unaided is no less consequential, only less auditable. Questions of rule authorship, along with whether users should interrogate, override, contest, or escalate cues, fall within a broader governance layer distinct from the present decision-level framework.

\section{Conclusion}
Meaningful human oversight of AI-assisted decisions requires visibility into how uncertainty conditions compose into a response, not merely observing the uncertainty itself. 
The framework and prototype remain preliminary, and empirical evaluation under realistic workloads is an important next step. Questions of user authority to interrogate, override, contest, or escalate cues likewise point toward a broader governance layer beyond the present work. Nevertheless, the central contribution remains: making the path from detected uncertainty to oversight response visible, rather than leaving that composition implicit within the system or the user.

\acknowledgments{The authors wish to thank Ebenezer Awotoro for discussions that led to the improvement of the paper}

\bibliographystyle{abbrv-doi}

\bibliography{template}
\end{document}